\documentclass[
floatfix,citeautoscript]{revtex4-2}
\usepackage{graphicx,psfrag,color}
\usepackage{slashed,amssymb}
\def\no{\noindent}
\def\bc{\begin{center}}
\def\ec{\end{center}}
\def\vs{\vskip0.5cm}
\def\beq{\begin{equation}}
\def\eeq{\end{equation}}

\def\bc{{\bf c}}

\def\bk{{\bf k}}
\def\ba{{\bf a}}
\def\br{{\bf r}}

\def\bv{{\bf v}}

\def\text{{\rm}} 

\begin{document}

\title{
Relation between winding numbers and energy dispersions
}
\author{Quancheng Liu$^{1}$ and Klaus Ziegler$^{2,3}$\\
$^1$School of Physics, State Key Laboratory of Crystal Materials,\\
Shandong University, Jinan 250100, China\\
$^2$Institut f\"ur Physik, Universit\"at Augsburg,
D-86135 Augsburg, Germany\\
$^{3}$Physics Department, New York City College of Technology,\\
The City University of New York, Brooklyn, NY 11201, USA
}
\date{\today}

\maketitle
\no
Abstract:

\no
Two-band Hamiltonians provide a typical description of topological band structures, 
in which the eigenfunctions can be characterized by a 
winding number that defines an integer topological invariant. This winding number is
quantized 
and protected against continuous deformations of the Hamiltonian. Here we show that 
the Bloch vector and its winding number can be directly related to the gradient of 
the energy dispersion. 
Since the energy gradient is proportional to the group velocity, our result establishes an
experimentally accessible correspondence between the Bloch vector field and 
angle-resolved photoemission spectroscopy measurements. We discuss a 
mapping between the gradient of the energy dispersion and the Bloch vector.
This implies a direct and measurable relation between two-band Hamiltonians and their
underlying topological structures.

\vs

In two-dimensional materials topological properties can play a crucial role. Their discovery
was very successful in explaining the quantum Hall effect~\cite{klitzing80}, since 
the Hall conductivity is a 
topological invariant~\cite{thouless82,haldane88,yakovenko90,hatsugai93}. 
Topological properties are also essential for the quantum spin Hall 
effect~\cite{kane05a,kane05b,bernevig06,koenig07}.
More recent works on artificial materials (e.g., Floquet dynamics on systems of quantum 
wells and waveguides) 
have focused on various topological properties~\cite{lindner11,rudner13,segev13,liu25}. 
This is often motivated by the existence of topological invariants 
such as the Berry phase or Chern numbers. In contrast to the quantum Hall effect, in these
systems the topological invariants are not always directly 
accessible and are more difficult to detect~\cite{segev13} or to 
calculate~\cite{rudner13,fukui05,fehske17}. Therefore, it might 
be also of practical interest to probe topological properties with tomographic methods.
In the following we propose to extend the study of invariants
to the analysis of trajectories on the Bloch sphere and their winding numbers. 
Our quantum systems are described by a Hamiltonian of the simple form 
$\vec{\chi}\cdot\vec{\sigma}$ with the vector of Pauli matrices $\vec{\sigma}$, introduced
for the quantum Hall effect without Landau levels~\cite{haldane88} and discussed
subsequently in numerous papers~\cite{yakovenko90,zhang06,fradkin,moore08}.

In a typical study only the eigenfunctions, but not the eigenvalues, reflect the
topological properties of 
a Hamiltonian~\cite{yakovenko90,zhang06,fradkin,moore10,keimer17}, 
such as the winding number of the eigenfunctions or the Berry phase. 
In this Letter we will argue that also the eigenvalues of a translation-invariant 
Hamiltonian can be used to calculate the winding number. 
This offers an approach to probe topological properties within conventional
experimental techniques, such as Angle-Resolved Photoemission Spectroscopy (ARPES).
For this purpose we consider a translation-invariant 
Hamiltonian on a physical space where the latter
is either two dimensional with real coordinates $\br$ or a corresponding 
Fourier space with wave vector $\bk$. Edges can be added subsequently and treated 
either through the topological bulk-edge correspondence~\cite{Graf2013,shapiro20} 
or through the analytic bulk-edge connection (ABEC)~\cite{ziegler2501}. 
Moreover, we assume that $H$ acts on a two-component spinor 
space (``two-band model''), such that its eigenfunctions $\{\Psi_k\}$ are
two-component spinors.

The main idea is based on 
an intimate connection between the energy eigenvalues and the Bloch vector (BV),
which can be seen in the change of the eigenvalue $E_k=(\Psi_k\cdot H\Psi_k)$ under the
change of the Hamiltonian $H$. $\Psi_k$ is the normalized eigenfunction $\Psi_k(\br)$
and the brackets represent the spatial integration, where
$\br$ are the coordinates on the physical space.
In the case of translation invariance the eigenfunctions are plane waves or 
in the case of rotational invariance they are spherical waves. 
The latter is applicable to a circular disk-like geometry~\cite{Ziegler:18,ziegler24}, 
but this case will not be pursued subsequently. 
In general, an infinitesimal change of $E_k$ with respect to a parameter $x$ reads
\beq
\label{relation00}
\partial_x E_k=(\Psi_k\cdot(\partial_xH)\Psi_k)
,
\eeq
since $(\Psi_k\cdot H(\partial_x\Psi_k))+((\partial_x\Psi_k)\cdot H\Psi_k)
=E_k\partial_x(\Psi_k\cdot\Psi_k))=0$ due to $(\Psi_k\cdot\Psi_k)=1$.
For the operator-valued Hamiltonian $H=\vec{\chi}\cdot\vec{\sigma}$,
we assume that $\vec{\chi}\Psi_k=\vec{h}\Psi_k$ with a number-valued 
vector $\vec{h}(\bk)$~\cite{haldane88,yakovenko90,zhang06},
where the specific choice of the eigenfunctions $\psi_k$ is dictated by geometry.
Then a simple relation between energy eigenvalues and the components of the BV
$\vec{s}_k=(\Psi_k\cdot\vec{\sigma}\Psi_k)$ emerges as
\beq
\label{relation0}
\nabla_{\vec{h}}E_k=\vec{s}_k
.
\eeq
On the other hand, the eigenvalues of the $2\times2$ matrix 
$H=\vec{h}\cdot\vec{\sigma}$ are $E_k=\pm h$, 
such that Eq. (\ref{relation0}) implies for the BV
\beq
\label{bloch1}
\vec{s}_k
=\pm\frac{1}{h}\vec{h}
.
\eeq
Thus, the BV field $\vec{s}_k$ is the result of a map from the physical space 
to the Bloch sphere $S^2$. In particular, a closed trajectory in the physical space 
results in a closed trajectory on the Bloch sphere, which is either
a single point, an open line, a single loop or a multiple loop with an 
integer winding number more than 1 or less than -1. Hence the winding numbers yield 
a classification of different maps $\vec{h}\to\vec{s}_k$.
Moreover, relation (\ref{bloch1}), which we derived from Eqs.
(\ref{relation00}) and (\ref{relation0}), enables us to recover the normalized vector 
$\vec{h}$ of the Hamiltonian from $\vec{s}_k$, 
while relation (\ref{relation0}) recovers the
gradient of the energy dispersion from $\vec{s}_k$.
In other words, for a given $\vec{h}$ we obtain directly the BV field from these relations,
without an explicit calculation of its eigenfunctions.
Thus, the mapping provides important information such as the topologically protected 
winding number. 

{\it Examples:}
To illustrate the usefulness of the relations (\ref{relation0}) and (\ref{bloch1}) in 
terms of a winding-number classification we consider some examples.
The first example is a two-band Laplacian Hamiltonian with mass $m$ and 
$\vec{h}(k)=(k^2,0,m)^T$, which represents a semi-conducting system with 
time-reversal symmetry. 
Its eigenvalue reads $E_k=\pm\sqrt{k^4+m^2}$ and its BV has only North-South
trajectories. Thus, it has the winding number $n_w=0$. 
The second example is the 2D Dirac Hamiltonian in Fourier space with
$\vec{h}(\bk)=(k_x,k_y,m)^T$ and energy dispersion $E_k=\pm\sqrt{k^2+m^2}$.
Replacing the Dirac mass $m$ by a third component of the wave vector $k_z$ we obtain 
the 3D Weyl
Hamiltonian~\cite{armitage18,burkov18}, where the BV of the former is the map 
$T^2\to S^2$ and the BV of the latter $T^3\to S^2$. These Hamiltonians have a 
winding number $n_w=1$.

\begin{figure}[t]
   \centering
a)
\includegraphics[width=7cm,height=6cm]{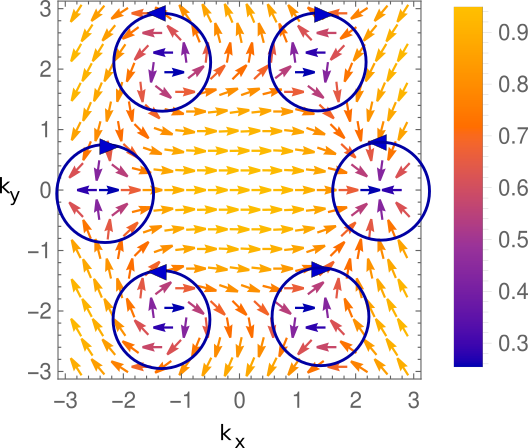}
b)
\includegraphics[width=6.5cm,height=7cm]{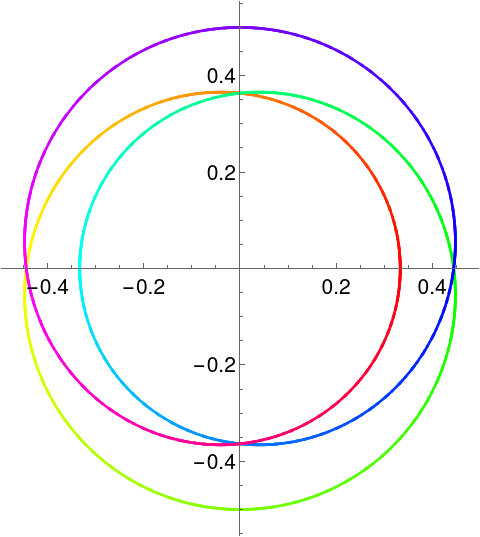}
\caption{
2D projection of the BV field: a) The BV field of a tight-binding Hamiltonian 
on a honeycomb lattice has several nodes. The marked loops around these nodes
have the winding numbers $\pm 1$, indicated by the clockwise or counter-clockwise winding,
respectively. 
The actual  length of the $s_1$-$s_2$ projected BV 
is color encoded. b) The winding trajectory of the BV of the BdG Hamiltonian 
with $w=n_w=3$. The continuous color change indicates the evolution of the trajectory 
with $t$ for $0\le t<2\pi$.
}    
\label{fig:pi_flux}
\end{figure}

A related lattice version of the 2D Dirac Hamiltonian is the tight-binding Hamiltonian on 
the honeycomb lattice with
\beq
\vec{h}(\bk)=(d',d'',m)^T
,\ \
d'=\sum_\mu\cos(\ba_\mu\cdot\bk)
,\ \
d''=\sum_\mu\sin(\ba_\mu\cdot\bk)
\eeq
with $\ba_1=a(0,-1)$ and  $\ba_{2,3}=a(\pm\sqrt{3},1)/2$ for a lattice constant $a$.
This Hamiltonian is time-reversal invariant, in contrast to the 2D Dirac or 3D Weyl
Hamiltonians.
In the lattice case the trajectory of the BV of a circle in $\bk$ space is more complex.
In particular, its winding number is not robust for all trajectories and, therefore, cannot be
associated with an invariant, as visualized in Fig.\ref{fig:pi_flux}a): it is $\pm1$ only near 
the nodes. This property is often used to obtain an integer Hall conductivity in low-energy
approximation~\cite{haldane88,fradkin}, where only small deviations from the
nodes are considered.
However, the integral over the entire Brillouin zone (i.e. the Chern number)
vanishes here due to the time-reversal invariance, as for the two-band Laplacian 
Hamiltonian. Nevertheless, the trajectories on the Bloch sphere are quite different
for both models, reflecting different topological properties.
All these Hamiltonians have been discussed extensively in the existing 
literature, especially in the low-energy
projection~\cite{haldane88,yakovenko90,zhang06,fradkin,moore08}.

The third example for $\vec{h}\cdot\vec{\sigma}$
is the Bogoliubov de Gennes (BdG) Hamiltonian that is qualitatively different,
since, in contrast to the other examples, the BV field is now hosted in a
four-dimensional space with coordinates $\bk$ and $\br$ such that
the BV is the result of the mapping $T^2\times T^2\to S^2$.
A non-uniform order parameter $\Delta(\br)=\Delta'(\br)+i\Delta''(\br)$ 
describes the Josephson effect~\cite{josephson62} and $\vec{h}$
is defined as
\beq
\vec{h}(\br,\bk)=(\Delta'(\br),\Delta''(\br),h_3(\bk))^T
\eeq
with energy dispersion $E_k=\pm\sqrt{|\Delta|^2+h_3(\bk)}$.
The order parameter reduces the translation invariance of the operator $\chi_3$. 
For example, if it is is periodic on the torus of length $L$ in $x$ direction with periodicity 
$w$, i.e., $\Delta=|\Delta|\exp(2\pi iwx/L)$, the winding number of the BV is $n_w=w$.
Thus, we can control the winding by the choice of the periodic order 
parameter and create a sequence of topological transitions~\cite{ziegler2510}.
Parametrizing a trajectory on $T^2\times T^2$  as $h_3=2+\cos^2 t$ and $x=t$ 
for $0\le t<2\pi$, with $w=3$ the trajectory on the Bloch sphere has winding 
number $n_w=3$, as visualized in Fig. \ref{fig:pi_flux}b).

{\it Physical interpretation of the BV:}
It is well known that $\vec{h}/h$ of the Hamiltonian $\vec{h}\cdot\vec{\sigma}$
is directly linked to the Berry curvature~\cite{yakovenko90,fradkin,moore08}.
This can be used to obtain the topological charge (or Chern number) as the integral over 
the Brillouin zone (BZ)
\beq
\label{chern_number}
C_\pm
=\pm\frac{1}{4\pi}\int_{\rm BZ}
\vec{s}\cdot(\partial_{k_x}\vec{s}\times\partial_{k_y}\vec{s})d^2\bk
,
\eeq
which has been linked to the Hall conductivity~\cite{yakovenko90,zhang06}.
With Eq. (\ref{relation0}) we can relate $C_\pm$ to the gradient of the energy dispersion now.   
The BV field provides more information than this integral though. In particular,
while the integral vanishes or gives an integer, the BV field provides
specific trajectories on the Bloch sphere in subspaces of the BZ, which characterize 
local properties, for instance, a winding number near spectral nodes.
Thus, it is suited for a tomographic analysis of complex quantum systems. 

Since the BV field is a gradient field with respect to $\vec{h}$, it is free of vortices.
However, it also depend on the physical space coordinates, either in real or 
Fourier space.
For instance, experimentally we may observe the dispersion as a function of the
wave vector $\bk$. To study the group velocity $\hbar \bv_\bk=\nabla_\bk E_k$ 
in ARPES~\cite{moore10},
we may consider the gradient of the dispersion as
\beq
\label{gradient_field0}
\nabla_\bk E_k
=\sum_\nu\frac{\partial E_k}{\partial h_\nu}\nabla_\bk h_\nu
=\vec{s}_k\cdot\nabla_\bk\vec{h}
,
\eeq
while a direct calculation yields
$\nabla_\bk E_k
=\pm\nabla_\bk h
$
due to $E_k=\pm h$. For the 2D Dirac and 3D Weyl Hamiltonian, where we have a linear
dispersion, the gradient of the
dispersion reproduces the BV field. In contrast, the BdG Hamiltonian leads to a gradient of the 
dispersion whose third component does not agree to the third component of the BV.

Besides the gradient of the energy dispersion in Eq. (\ref{relation0}), another interesting
analysis of the BV is associated with expectation values of the spinor components.
This can be understood
when we define for two spinor components $\psi_1$ and $\psi_2$ of 
$\Psi_k=(\psi_1,\psi_2)^T$ the following wave functions
\beq
\label{superpositions}
\psi_\pm:=\psi_1\pm\psi_2
,\ \ \psi_\pm':=\psi_1\pm i\psi_2
\eeq
and the corresponding probabilities $P_j=(\psi_j\cdot\Psi_j)$ as well as
\beq
P_\pm:=(\psi_\pm\cdot\psi_\pm)
, \ \ 
P_\pm':=(\psi_\pm'\cdot\psi_\pm')
.
\eeq
Then a straightforward calculation yields
\beq
\vec{s}_k
=\pmatrix{
P_+-P_- \cr
P_-'-P_+'\cr
P_1-P_2 \cr
}
,
\eeq
where the third component describes the polarization $2P_1-1$ of the two bands, 
since $P_1+P_2=1$. The other two components represent the polarization of the
super-positioned wave functions in Eq. (\ref{superpositions}).
This form of $\vec{s}_k$ is also known as the Stokes vector 
in classical optics and light scattering~\cite{hulst81,mishchenko06}.
To measure these probabilities we must have access to both spinor components
individually, which might be possible with Spin-ARPES~\cite{lashell96,hu06}.

{\it Edge modes:}
So far,
our discussion has focused on a compact physical space without edges,
as presented by a torus. For an additional edge, obtained by drilling a hole or cutting the torus, 
the properties of the edge modes can be derived via the ABEC
through an analytic continuation $k_j\to ic_j$ under the condition that the eigenvalues 
are real~\cite{ziegler2501}.
We note that this approach is based on the eigenfunctions in contrast to the topological 
bulk-edge correspondence~\cite{Graf2013,shapiro20}, which is based on the relation of 
the topological indices of the bulk (typically the Chern numbers) and the number of 
edge states. The analytic continuation results in a BV field, in which
an imaginary wave number represents a bound state in the direction perpendicular 
to the edge. The ABEC has some similarity with the concept of the Generalized
Brillouin Zone (GBZ) for non-Hermitian systems. In the latter, the eigenvalues are
imaginary though~\cite{yokomiz021}, while in the ABEC all eigenvalues are real.

For the BdG Hamiltonian, an edge along the $x$-axis creates edge modes in the
$y$-direction and a BV field $\vec{s}_k(\br,k_x,ic_y)$ with real $k_x$ and $c_y$. 
Boundary conditions on the edge fixes $c_y$ but leave $\br$ 
and $k_x$ as quasi-continuous variables. Thus, the BV
can be parametrized by the three-dimensional $x$-$y$-$k_x$ coordinates, where loops
in this space create winding trajectories on the Bloch sphere as those visualized in 
Fig. \ref{fig:pi_flux}b).  

{\it Conclusions:}
We have derived a direct relation between the $SU(2)$ Hamiltonian 
$\vec{\chi}\cdot\vec{\sigma}$
and the associated BV field, as given in Eq. (\ref{bloch1}). A central result is that the BV 
can be obtained directly from the gradient of the energy dispersion Eq. (\ref{relation0}), 
such that the band structure alone suffices to reconstruct the BV. While the BV yields 
the Chern number via Eq. (\ref{chern_number}), it provides a more general 
characterization that does not rely on integration 
over a predefined region of momentum space, allowing for flexible boundary conditions and
local topological analysis. Our approach establishes a direct identification between the BV, 
the energy dispersion, and the vector $\vec{h}$
of the $SU(2)$ Hamiltonian. Experimentally, the relation between the BV and the energy
gradient implies that angle-resolved photoemission spectroscopy can access the BV field
through measurements of the group velocity Eq. (\ref{gradient_field0}). Finally, 
we have seen that Hamiltonians of the form $\vec{\chi}\cdot\vec{\sigma}$
can be classified by the winding numbers of the BV field, providing a topological
characterization based on invariant subspaces rather than global integrals.

\end{document}